\documentclass[%
 preprint,
superscriptaddress,
 amsmath,amssymb,
 aps,
]{revtex4-2}

\usepackage{graphicx}
\usepackage{dcolumn}
\usepackage{bm}

\begin{document}

\preprint{APS/123-QED}

\title{Optical manipulation of Rashba-split 2-Dimensional Electron Gas}

\newcommand{\ubc}{Department of Physics \& Astronomy, University of British Columbia, Vancouver, BC V6T 1Z1, Canada }
\newcommand{\qmi}{Quantum Matter Institute, University of British Columbia, Vancouver, BC V6T 1Z4, Canada}
\newcommand{\mpi}{Max Planck Institute for Chemical Physics of Solids, Dresden, Germany}
\newcommand{\aau}{Department of Physics and Astronomy, Interdisciplinary Nanoscience Center, Aarhus University, 8000 Aarhus C, Denmark}
\newcommand{\aauc}{Department of Chemistry, Aarhus University, 8000 Aarhus C, Denmark}
\newcommand{\mon}{Centre \'{E}nergie Mat\'{e}riaux T\'{e}l\'{e}communications, Institut National de la Recherche Scientifique, Varennes, Qu\'{e}bec J3X 1S2, Canada}

\author{M.\,Michiardi}
\email{mmichiardi@phas.ubc.ca}
\affiliation{\qmi}
\affiliation{\ubc}
\affiliation{\mpi}
\author{F.\,Boschini}
\affiliation{\qmi}
\affiliation{\ubc}
\affiliation{\mon}
\author{H.-H.\,Kung}
\affiliation{\qmi}
\affiliation{\ubc}
\author{M.\,X.\,Na}
\affiliation{\qmi}
\affiliation{\ubc}
\author{S.\,K.\,Y.\,Dufresne}
\affiliation{\qmi}
\affiliation{\ubc}
\author{A.\,Currie}
\affiliation{\qmi}
\affiliation{\ubc}
\author{G.\,Levy}
\affiliation{\qmi}
\affiliation{\ubc}
\author{S.\,Zhdanovich}
\affiliation{\qmi}
\affiliation{\ubc}
\author{A.\,K.\,Mills}
\affiliation{\qmi}
\affiliation{\ubc}
\author{D.\,J.\,Jones}
\affiliation{\qmi}
\affiliation{\ubc}
\author{J.\,L.\,Mi}
\affiliation{\aauc}
\author{B.\,B.\,Iversen}
\affiliation{\aauc}
\author{Ph.\,Hofmann}
\affiliation{\aau}
\author{A.\,Damascelli}
\email{damascelli@physics.ubc.ca}
\affiliation{\qmi}
\affiliation{\ubc}

\date{\today}

\maketitle

\textbf{Abstract:}
In spintronics, the two main approaches to actively control the electrons’ spin involve static magnetic or electric fields. An alternative avenue relies on the use of optical fields to generate spin currents, which can bolster spin-device performance, allowing for faster and more efficient logic. To date, research has mainly focused on the optical injection of spin currents through the photogalvanic effect, and little is known about the direct optical control of the intrinsic spin-splitting. To explore the optical manipulation of a material’s spin properties, we consider the Rashba effect. Using time- and angle-resolved photoemission spectroscopy (TR-ARPES), we demonstrate that an optical excitation can tune the Rashba-induced spin splitting of a two-dimensional electron gas at the surface of Bi$_{2}$Se$_{3}$. We establish that light-induced photovoltage and charge carrier redistribution – which in concert modulate the Rashba spin-orbit coupling strength on a sub-picosecond timescale – can offer an unprecedented platform for achieving optically-driven spin logic devices.

\begin{figure*}
    \centering
    \includegraphics[width=0.75\textwidth]{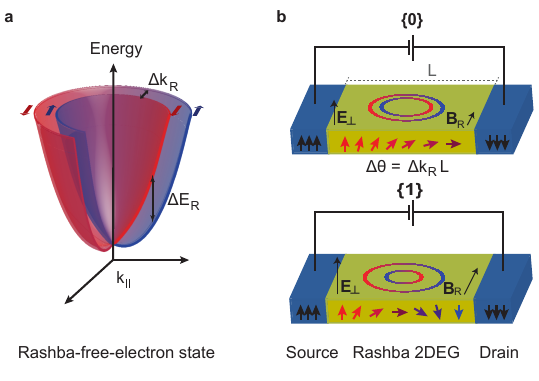}
    \linespread{1.3}
    \caption{\textbf{Rashba spin-orbit coupling in two-dimensional electron gas.} \textbf{a}: Rashba spin-orbit-coupling (RSOC) splits a free electron state into two subbands carrying opposite spin texture (red and blue). The splitting of the free electron state in both energy ($\Delta E_{R}$) and momentum ($\Delta k_{R}$) is proportional to the RSOC strength $\alpha_{R}$, which is tunable with an electric field $E_{\perp}$. This Rashba splitting locks the electron's spin to its momentum. \textbf{b}: Fundamental design of a spin field-effect transistor (spinFET) in which spin-polarized electrons are injected from a source into a Rashba 2DEG and collected with a ferromagnetic drain. Due to the momentum-dependent splitting of Rashba 2DEGs, charges traversing from source to drain feel an effective magnetic field, $B_{R}$, proportional to $\alpha_{R}$, perpendicular to their direction of motion, causing their spin to precess. The spin polarization of carriers changes by the angle $\Delta\Theta = \Delta k_{R}\cdot L$, where L is the length of the 2DEG. Modulating $\Delta k_{R}$ -- conventionally via an electric field -- switches the spinFET between a state of high \{0\} and low \{1\} resistance.}
    \label{fig:fig1}
\end{figure*}

\textbf{Introduction:}
Spintronics has the potential to deliver computational devices that are less volatile, faster, and more energy efficient with respect to their electronic counterparts \cite{Oestreich:2001}. However, the need to control the spin degree of freedom in a fast and efficient manner is challenging, as the field required to flip the electron's spin in magnetic materials is often prohibitively high. Spin-orbit coupling (SOC) effects, such as the Rashba effect, allow the formation of spin-polarized electron states without a magnetic moment, thereby circumventing this limitation. In particular, the Rashba effect manifests as a broken spin degeneracy at semiconductor interfaces, resulting in quasi-particle bands of opposite spin texture that are offset in momentum \cite{Bychkov:1984, Rashba:2006}. The Rashba effect has long been a staple in the field of spintronics owing to its superior tunability, which allows the observation of fully spin-dependent phenomena, such as the spin-Hall effect, spin-charge conversion, and spin-torque in semiconductor devices \cite{Bercioux:2015, Soumyanarayanan:2016}. An example of a Rashba-split quasi-free electron state with effective mass $m^{*}$ is shown in Fig.\,\ref{fig:fig1}\textbf{a}. To the first order, its dispersion relation is given by: 
\begin{equation}
E=\frac{\hbar ^{2}k^{2}}{2m^{*}}\pm \alpha_{R} k.
\label{eq:rashba}
\end{equation}
Here, the parameter $\alpha_{R}$ is the strength of the Rashba SOC (RSOC) in the system, and it depends on the atomic SOC as well as the electric field perpendicular to the surface (\textbf{E}$_{\perp}$). Experimentally, $\alpha_{R}$ can be extracted from the detailed dispersion of the spin-split subbands: the energy splitting of the subbands is given by $\Delta E_{R} = 2\alpha_{R}k$, and can be seen as a momentum-dependent Zeeman splitting caused by the pseudo-magnetic field -- or Rashba field -- $\textbf{B}_{R}\propto\textbf{k}\times\textbf{E}_{\perp}$; correspondingly, the momentum splitting is given by $\Delta k_{R} = 2\alpha_{R}m^{*}/\hbar^{2}$. 
 
We illustrate the inner workings of these parameters with the paradigmatic example of the spin field-effect transistor (spinFET), depicted in Fig.\,\ref{fig:fig1}\textbf{b}. In the pioneering concept of Datta and Das \cite{Datta:1990}, a Rashba-split two-dimensional electron gas (2DEG) in a channel of length $L$ is sandwiched between two spin-polarized leads. As electrons transit the 2DEG in the direction perpendicular to the Rashba field  $\textbf{B}_{R}$, their spin precesses, acquiring a phase $\Delta\Theta = \Delta k_{R}L$ (assuming the chemical potential lies above the bands' degeneracy point). Switching between the {0}/{1} logic operation – corresponding to the low/high resistance state in the device – is achieved by tuning $\Delta k_{R}$ such that the electron spin at $L$ aligns to that of the drain lead. As $\Delta k_{R}$ is proportional to $\alpha_{R}$,  the operation of such a device relies primarily on the possibility to tune the RSOC, typically realized by gating the 2DEG \cite{Chuang:2015}.

In spintronic devices such as the spinFET, the prospect to replace the gate with an optical field prompts the development of even faster and more efficient hybrid opto-spintronics. To this end, previous works have demonstrated the generation of spin-polarized currents in Rashba and topological states through the photogalvanic effect, as well as the ultrafast switching of spin orientation in antiferromagnets \cite{Yuan:2014, Liu:2020, Kimel:2004, Nova:2016, McIver:2009, Wang:2007, Zhou:2006}, however little is known about the direct optical control of the intrinsic spin splitting \cite{Sheremet:2016, Cheng:2017}. Here, we show that light can change the RSOC strength, effectively manipulating the Rashba spin-transport properties on an engineered 2DEG. The proposed mechanism is as follows: in the presence of a band-bending surface potential, an above-gap optical excitation drives a charge redistribution along the axis perpendicular to the surface. This charge redistribution creates an ultrafast photovoltage, which then reliably alters the RSOC strength ($\alpha_R$) of the 2DEG system on a sub-picosecond timescale. We employ time- and angle-resolved photoemission spectroscopy (TR-ARPES) to track the evolution of the RSOC strength through the dispersion of the Rashba 2DEGs. By directly measuring $\Delta k_R$ and $\Delta E_R$ as a function of pump-probe delay, we unambiguously extract the evolution of $\alpha_R$.

\textbf{Results:}

\begin{figure*}
    \centering
    \includegraphics[width=0.6\textwidth]{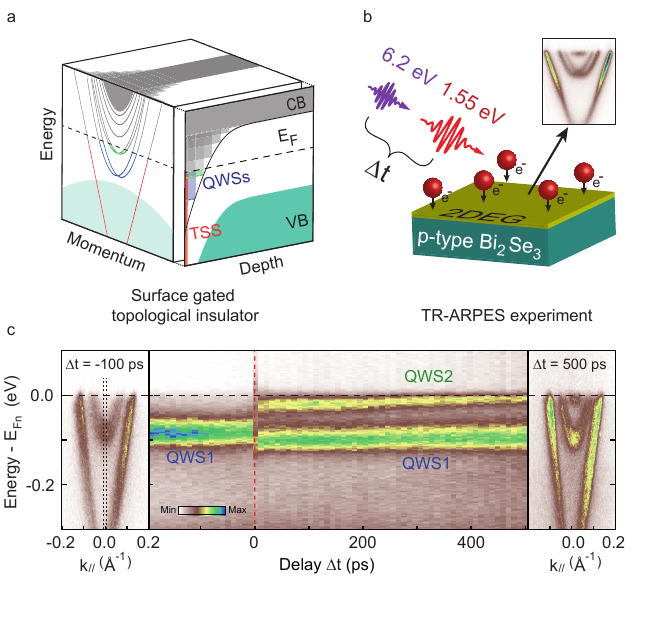}
    \linespread{1.3}\selectfont{}
    \caption{
    \textbf{TR-ARPES of surface-gated topological insulators.} 
    \textbf{a}: Representation of the surface and bulk electronic structure in a surface-gated topological insulator as a function of momentum, energy, and distance from the surface (the side view displays the momentum integrated projection of the band structure, where CB and VB are the conduction and valence bands). Two-dimensional electron gases (2DEGs) taking the form of spatially confined quantum well states (QWSs) are created by a sufficiently large positive bias applied to the surface. The dispersion, Rashba-splitting, and spatial extent of the 2DEGs depend on the detailed shape of the band bending. Here, the band bending pushes the two lowest QWSs (blue and green) below the Fermi energy.
    \textbf{b}: TR-ARPES experiment on p-type Bi$_{2}$Se$_{3}$; the cleaved sample is gated \textit{in situ} by alkali atom deposition. A near-infrared (1.55\,eV) ''pump" pulse perturbs the system, and a UV (6.2\,eV) pulse is used to probe the electronic structure by ARPES. The time delay ($\Delta t$) between pump and probe pulses is varied to resolve the electron dynamics. \textbf{c}: Temporal evolution of the QWSs in p-doped Bi$_{2}$Se$_{3}$ plotted relative to the electron quasi-Fermi level E$_{Fn}$. The left panel shows the ARPES spectra of all surface states before pump arrival (-100\,ps); in the center panel photoemission intensity integrated around the Brillouin zone center (black dashed lines) is shown as a function of time (pump and probe are overlapped at time zero, red dashed line). We observe that a second QWS emerges after the pump excitation; the right panel shows the dispersion at 500\,ps, characterized by two partially populated QWSs.}
    \label{fig:fig2}
\end{figure*}

Among the materials that can host Rashba-split 2DEGs, bismuth-based topological insulators (TI) are an ideal platform: 2DEGs can be induced on the surface of TIs  by applying a positive surface bias or chemical gating \cite{Bianchi:2010,Bahramy:2012, Zhu:2011,Michiardi:2015}. The combination of the strong atomic SOC in TIs with surface gating generates a substantial Rashba effect in the 2DEGs, allowing one to finely resolve the spin splitting. 
In an ideal TI, only the topological surface state (TSS) -- recognizable by its linear dispersion across the bandgap -- crosses the Fermi level ($E_F$), and all charge carriers belong to the TSS \cite{Barreto:2013}. As represented in Fig.\,\ref{fig:fig2}\textbf{a}, the application of a sufficient positive bias at the surface induces a strong band bending, leading to the creation of 2DEGs in the form of surface confined quantum well states (QWSs). While the TSS wavefunction extends only within a few layers from the surface and does not depend on the shape of the surface potential, the wavefunction of the QWS does, and extends comparatively deeper into the bulk \cite{Bahramy:2012}. The difference in spatial extent between the TSS and QWS wavefunctions allows us to extract the behavior specific to QWSs, as opposed to the behavior of surface states in general. 

Our experimental approach is depicted in Fig.\,\ref{fig:fig2}\textbf{b}. We choose p-doped Bi$_{2}$Se$_{3}$ to host QWSs, as the hole doping provides a lower bulk conductivity in this material. The 2DEGs are prepared by depositing a controlled amount of alkali atoms on the surface, leading to a population of conduction band-derived states that are spin-split by the Rashba effect. Increasing the concentration of deposited atoms is analogous to raising the surface bias, which introduces a higher surface charge density and stronger band bending. The system is then optically excited with a near-infrared (1.55\,eV) ''pump" pulse and its response is probed by photoemission using a UV (6.2\,eV) pulse at variable time delay $\Delta$t \cite{Perfetti:2006,Freericks:2008}. 
The result of such a TR-ARPES experiment is summarized in Fig.\,\ref{fig:fig2}\textbf{c} over a long range of delays. The left and right panels show the ARPES spectra at negative delay (-100\,ps), and 500\,ps after the pump arrival, respectively. The central panel presents the evolution of the states at the Brillouin zone center (black dashed lines in the left panel). Before the pump arrival (-100\,ps), the system is in equilibrium; the Fermi level is crossed by the linear topological surface state (TSS) and a single parabolic band, nominally the first quantum well state (QWS1). Here, the Rashba-splitting is just barely discernible, owing to the moderate chemical gating. At zero-delay, electrons are optically excited into unoccupied states and subsequently decay into a quasi-thermalized state \cite{Sobota:2014, Hajlaoui:2012}. Remarkably, we see that QWS1 is pushed to lower energies after the excitation, and a second band becomes populated. This second band (shown also in the spectrum at 500\,ps) is in fact the second quantum well state (QWS2), which emerges following an increase in surface charge density. It is worth noting that the aforementioned photovoltage induced by the pump pulse also affect the kinetic energy of photoemitted electrons \cite{Ciocys:2019, Yang:2013, Tanaka:2012}. This manifests as a rigid shift of the ARPES spectra that can be accounted for by a simple subtraction; henceforth, we refer all energy scales to the electron quasi-Fermi level E$_{Fn}$, extracted by fitting a Fermi-Dirac distribution to the photoemission intensity around the TSS Fermi vector (details can be found in the supplementary information).

\begin{figure*}[!p]
    \centering
    \includegraphics[width=1.0\textwidth]{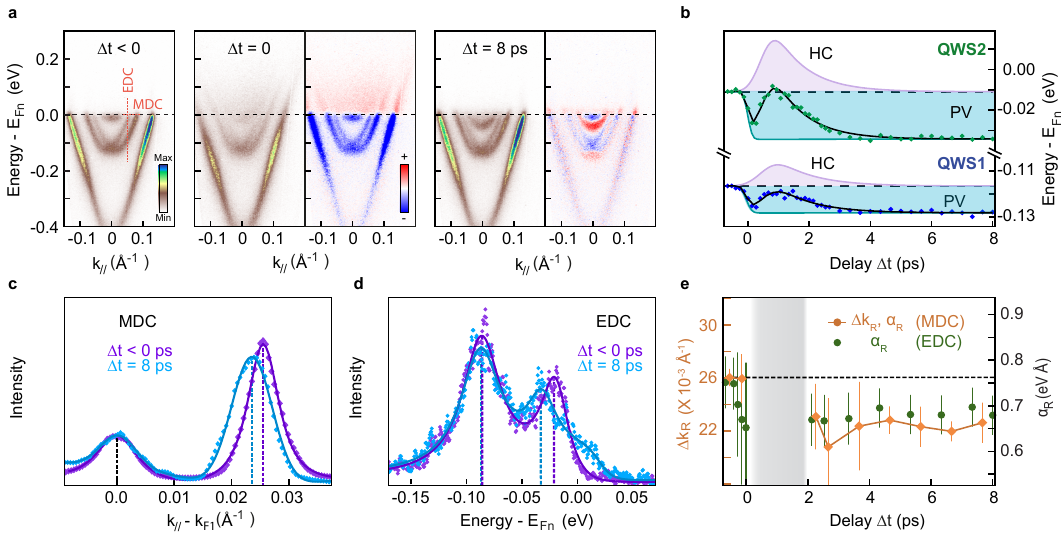}
    \linespread{1.0}\selectfont{}
    \caption{\textbf{Ultrafast response of Rashba QWSs to optical perturbation.} \textbf{a}: ARPES dispersion at negative time delay (before pump arrival), at time zero (pump and probe fully overlapped), and at 8\,ps. The latter two are accompanied by differential spectra obtained by subtracting the spectrum acquired at around -0.5\,ps; the blue (red) color is indicative of a pump-induced decrease (increase) of photoemission intensity.
    \textbf{b}: Temporal evolution of the energy minimum for QWS1 and QWS2  extracted from fitting the ARPES data at $k_{\parallel}=0$. The fit to the experimental data (solid black line) stems from two contributions, each consisting of a finite rise-time step function and an exponential decay. The positive contribution (purple) is defined as a hot-carriers (HC) driven  process, which is short-lived; the negative contribution (cyan) is the result of a photovoltage (PV) effect.
    \textbf{c}: Momentum distribution curves (MDC) profiles across the right branch of QWS1 (horizontal dashed line in \textbf{a}) at E$_{Fn}$ at equilibrium (purple) and after 8\,ps (blue) relative to the Fermi wave-vector of the inner branch $k_{F1}$; the solid lines are Voigt fits to the data. The momentum splitting $\Delta k_{R}$ is the distance between two peaks of the same MDC, and it is dynamically reduced with the pump excitation from (26\,$\pm$\,0.5)$\times 10^{-3}$ (at $\Delta t <$ 0) to (22.5\,$\pm$\,0.5)$\times 10^{-3}$\,\AA$^{-1}$ (at $\Delta t$ = 8\,ps).
    \textbf{d}: Energy distribution curves (EDC) profiles across the inner and outer branch of QWS1 at $k_{\parallel}$ = 0.05\,\AA$^{-1}$ (vertical dashed line in \textbf{a}) before pump arrival (purple) and after 8\,ps (blue). The optical excitation also induces the reduction of the energy splitting.
    EDC and MDC profiles in  \textbf{c} and \textbf{d} have been deconvolved by the energy resolution via Lucy-Richardson algorithm (Ref.\,\onlinecite{Yang:2008}) for better clarity. Both profiles are fitted using Voigt functions (solid lines).
    \textbf{e}: Temporal evolution of the RSOC strength $\alpha_{R}$ in QWS1 calculated from Eq.\,\ref{eq:rashba} and extracted by fitting the momentum (orange) and energy (green) splitting at several time delay; $\alpha_{R}$ is reduced by about 0.1\,eV\AA\, at 8\,ps with respect to equilibrium. The values of $\Delta k_{R}$ at E$_{Fn}$ are explicitly plotted against the left axes. At 0 $< \Delta t <$ 2 the signal is too low to convey reliable physical significance. All the values of energy and momentum splitting are obtained by fitting the raw data, and error bars are evaluated from statistical distribution within 95\% confidence. }
    \label{fig:fig3}
\end{figure*}

In pursuance of determining the impact of the optical excitation on the Rashba effect, we perform TR-ARPES on a sample with a higher concentration of deposited alkali atoms, so that energy and momentum splittings are better distinguished. The results of this experiment are shown in Fig.\,\ref{fig:fig3}. In panel \textbf{a}, the dispersion is shown for three pump-probe delays ($\Delta$t=-0.5, 0, and 8\,ps). We observe that both QWSs (parabolic bands) are populated before pump arrival, with energy minima at -114 and -13\,meV, and QWS1 exhibits a visible and strong Rashba splitting. Differential ARPES maps [$I(\mathbf{k},E,t)-I(\mathbf{k},E,-0.5$\,ps)] of the 0 and 8\,ps delays are also shown, highlighting the pump-induced modification of the QWSs. At time zero, the optical excitation creates an electron population (depletion) above (below) E$_{Fn}$, but shows no appreciable change in dispersion. However, at 8\,ps, while the TSS shows no significant change, both QWSs shift downwards in response to an increase in surface charge, similarly to what was reported in Fig.\,\ref{fig:fig2}(c). This is further emphasized in Fig\,\ref{fig:fig3}\textbf{b} where the time dependent energy shifts of the QWSs at Brillouin zone center are displayed. For both QWSs, the energy minimum shows a fluctuation at short timescales before eventually settling to lower energy. We fit the curves in Fig.\,\ref{fig:fig3}\textbf{b} with a phenomenological model that includes two exponentially decaying processes, shown in purple and cyan respectively (note that the latter curve appears flat because of a long decay time). The first process acts to increase the QWSs' energy, peaking at approximately 1.5\,ps after the pump excitation, and decaying within 3\,ps. The dynamics of this components follows the same temporal evolution of the electronic temperature in the system (shown in Supplementary Information), and the timescales are characteristic of the optically-driven electron population above the Fermi level in TIs \cite{Sobota:2014}; therefore, we attribute this process to an effect caused by the presence of hot carriers (HC) close to E$_{Fn}$. The effect of HC on the surface potential can be extremely complex, but it is likely that, the mobile hot carriers further screen the built-in electric field, causing the QWSs' to shift to higher energy.    
The second and more interesting process is a long-lasting shift of the QWSs to lower energy, that emerges as a consequence of an increase in the surface electron population and variation of the electrostatic environment. This process -- as we will show in detail -- is given by a photovoltage (PV) effect, and it is the central mechanism of this work. The effect of the PV arises within a few hundred femtoseconds and alters the energy and density of the QWSs over hundreds of picoseconds. 

As the PV alters the electrostatic environment at the surface, we expect it will have an impact on the Rashba effect as well. For a quantitative look at the momentum splitting, we plot in Fig.\,\ref{fig:fig3}\textbf{c} the momentum distribution curves (MDCs) of QWS1 at E$_{Fn}$, in equilibrium (before the pump arrival) and 8\,ps after the excitation. The MDCs span the two spin-polarized bands on the right-hand side of QWS1 (see red dashed line in Fig.\,\ref{fig:fig3}\textbf{a}) and are referenced to the Fermi momentum of the inner branch ($k_{F1}$). The MDC peak locations are indicated by dashed lines; in comparing the equilibrium (purple) and post-excitation (blue) MDCs, we observe that the momentum splitting of the carriers is reduced from (26.0\,$\pm$\,0.5)$\times 10^{-3}$ to (22.5\,$\pm$\,0.5)$\times 10^{-3}$\,\AA$^{-1}$. 
A similar result is observed for the energy splitting: in Fig.\,\ref{fig:fig3}\textbf{d}, we plot the energy distribution curves (EDCs), for the same two delays, along the cut shown in Fig.\,\ref{fig:fig3}\textbf{a}; we find that, whereas the outer branch of QWS1 maintains its position, the inner branch moves to lower energy, leading to a reduction of the energy splitting $\Delta E_{R}$ by approximately 13\,meV. The additional shoulder observed in the EDC at 8\,ps arises from QWS2, which also moves to significantly lower energy. The simultaneous reduction of both $\Delta k_{R}$ and $\Delta E_{R}$ is a clear indication of an optically driven change of the Rashba spin-orbit coupling strength $\alpha_{R}$, and excludes modifications of the electron dispersion and effective mass as relevant contributions.

The full temporal dynamics of the Rashba effect in QWS1 under optical excitation is given in Fig.\,\ref{fig:fig3}\textbf{e}, where the splitting in momentum $\Delta k_{R}$ (in orange), as well as the RSOC strength obtained following Eq.\,\ref{eq:rashba}, are shown on the left and right y-axis, respectively. We observe that $\Delta k_{R}$ decreases immediately after the excitation, and -- after less than 3\,ps -- is effectively reduced to a seemingly constant value. The change in the RSOC strength is about 15\%, decreasing from a value of 0.76\,$\pm$\,0.02 to 0.66\,$\pm$\,0.02\,eV\AA. Similar values of $\alpha_R$ are obtained by performing an analogous analysis on $\Delta E_{R}(k)$, plotted in green; here, the MDC and EDC analysis between 0 and 2\,ps could not be performed due to the presence of a highly non-thermal electronic distribution. Our data outline a scenario where an opportune optical pulse changes the RSOC strength, in a manner similar to a static electric field, on a picosecond time-scale.   

The observation of an increased surface electron density in concert with a decrease in RSOC is, however, nontrivial, as the two quantities typically increase/decrease correspondingly. Thus, a satisfactory explanation for the observed effect of the pump excitation, requires one to consider the detailed variation of the surface electric field in relation to the spatial electron distribution. To this end, we build a model to capture the salient aspects of the experimental results. We begin by calculating the band bending of the system at equilibrium, which can be described by a one-dimensional model in the out of plane direction $x$. The potential profile $V(x)$ is calculated by solving the Poisson equation within a modified Thomas-Fermi approximation \cite{Paasch:1982,King:2008}. The  binding energy and wavefunction of the QWSs is computed \textit{a posteriori} by solving the Schr\"odinger equation within the calculated $V(x)$. For our simulation, all material-specific parameters for the calculation are taken from Refs. \onlinecite{Analytis:2010,Martinez:2017,Nechaev:2013,Gao:2014}, and the surface potential $V_{0}$ is determined empirically by the shift of the TSS Dirac point induced by chemical gating (see Supplementary Information). 

\begin{figure*}
    \centering
    \includegraphics[width=1.0\textwidth]{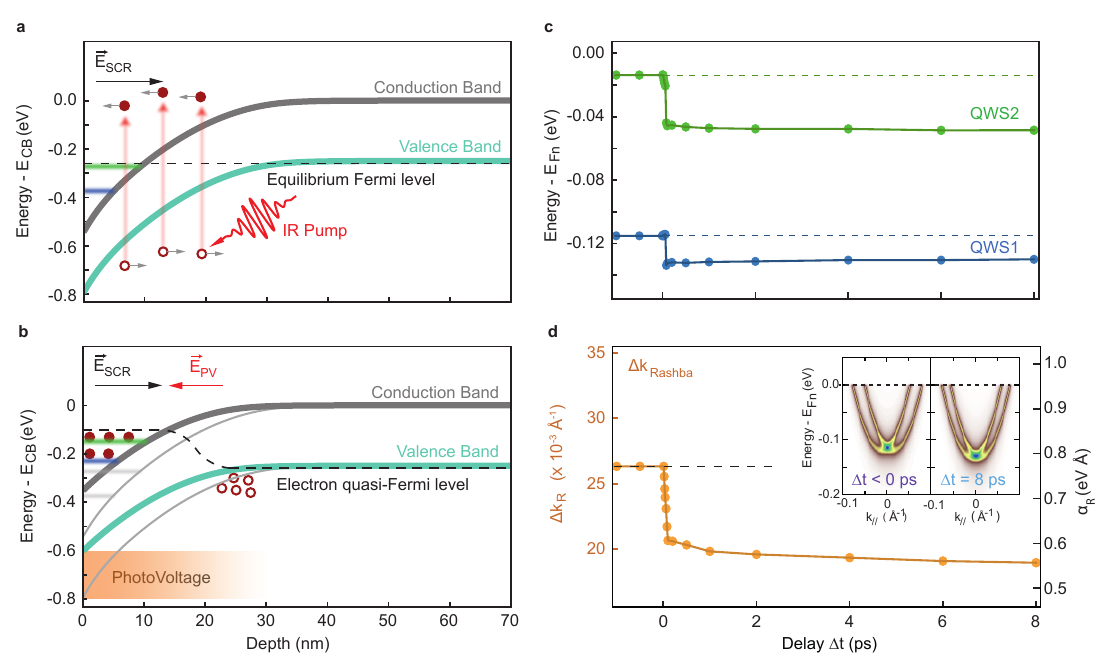}
    \linespread{1.3}\selectfont{}
    \caption{\textbf{Simulations of the band bending and quantum-well state dynamics.} \textbf{a}: Band-bending profile for a surface biased p-doped Bi$_{2}$Se$_{3}$; the surface boundary condition $V_0$ is given by the shift of the Dirac point after potassium evaporation, and the bending is calculated solving the Poisson equation within a modified Thomas-Fermi approximation. The energy minima of QWS1 and QWS2 are solutions of the Schr\"{o}dinger equation within the confining potential, and well reproduce experimental observations. At zero-delay, an optical excitation creates free electron-hole pairs that are swept apart by the built-in electric field of the space-charge region, E$_{SCR}$. \textbf{b}: After 8\,ps the charge separation between electrons and holes generates a photovoltage whose electric field (\textbf{E$_{PV}$}) opposes \textbf{E$_{SCR}$} and softens the band bending, causing the QWSs to shift upwards. Concomitantly, the increasing surface electron density shifts the E$_{Fn}$ upwards. \textbf{c}: The calculated energy minima of QWS1 and QWS2 with respect to E$_{Fn}$ are obtained from a time-dependent calculation of the band bending and QWSs' energy levels; the observed decrease at time zero is given by the change of the electrostatic environment and carrier redistribution across SCR. \textbf{d}: The simulated Rashba momentum splitting ($\Delta k_{R}$) at the Fermi level for QWS1; the Rashba strength $\alpha_{R}$ is given on the right axes. The photovoltage-induced reduction of the surface electric field in the system is responsible for the decrease of $\alpha_{R}$ and $\Delta k_{R}$. Inset: Simulated spectral functions of QWS1 before and 8~ps after the pump excitation, constructed from the results of the dynamical simulation. The simulated data confirms the two main observation from the TR-ARPES experiment, an increase in electron population accompanied by a decrease in the spin-splitting.}
    \label{fig:fig4}
\end{figure*}

We present the calculated equilibrium potential and QWSs' energy minima in Fig.\,\ref{fig:fig4}\textbf{a}. The space charge region (SCR) spans more than 30\,nm, and QWS1 and QWS2 are partially populated, replicating the experimental observations of Fig.\,\ref{fig:fig3}. Following an optical excitation across the band gap, the generated electron-hole pairs within the SCR are swept apart by the electric field \textbf{E$_{SCR}$}, which pushes the negative charges towards the surface and the positive charges towards the bulk \cite{Kronik:1999}. The electrons and holes become spatially separated over tens of nanometers, giving rise to a long-lasting photovoltage field \textbf{E}$_{PV}$ of the opposite sign to \textbf{E$_{SCR}$}. This effectively softens the band bending, pushing the surface potential V$_{0}$ to less negative values, as shown in Fig.\,\ref{fig:fig4}\textbf{b}. The new shallower surface potential drives both the QWSs and the $E_{Fn}$ to higher energy. To accommodate the surplus of surface electric charge, the $E_{Fn}$ shifts further upwards, resulting in the QWSs moving to more negative energies when plotted with respect to $E_{Fn}$, as seen in the TR-ARPES data (details in Supplementary Information).
It must be noted that, as small surface state-induced band bending is a common feature in semiconductors, surface PV has previously been observed in pristine TIs \cite{Yoshikawa:2019, SanchezBarriga:2017, Ciocys:2020}. However, while this is technologically relevant for TIs -- because it leads to spin-polarized diffusion currents -- the TSS only undergoes a rigid shift in energy under the surface PV. The 2DEGs, on the other hand, are much more sensitive to the shape and magnitude of the confining potential $V(x)$, and by extension, the PV.

The full temporal dynamics of the QWSs has been simulated by introducing a pair of effective photo-charges in the system, which approximates the collective charge motion via a center of mass approach \cite{MoraSero:2006}. At each time step, the charge distribution, electric field and E$_{Fn}$ are reevaluated. The time evolution of the QWSs' energy with respect to the E$_{Fn}$ -- shown in Fig.\,\ref{fig:fig4}\textbf{c} -- is in good qualitative agreement with experimental data. Both QWS1 and QWS2 fall to more negative values almost instantaneously at positive delays. Consistent with experimental observations, the energy shift of QWS2 is larger than that of QWS1. This is a manifestation of the shallower confining potential, which allows for a smaller energy difference between consecutive QWSs. 
Lastly, the RSOC strength and, by extension, Rashba-splitting for QWS1 are calculated at each step from Eq.\,\ref{eq:rashba}. The evolution of $\Delta k_{R}$ is also consistent with experimental findings (Fig.\,\ref{fig:fig4}\textbf{d}), confirming that the reduction in the Rashba strength is due to the PV-induced softening of the surface potential.

With the exception of the fluctuation at early delays given by the presence of hot carriers, our simple model succeeds in reproducing all salient features of the experimental data, proving that a photovoltage is responsible for the observed behavior. The small quantitative deviations between the simulated and measured PV effect can be attributed to the model simplicity and approximations, such as the omission of changes in the screening and dielectric properties of the material induced by the PV. More complex simulations of the PV effect such as those recently developed in ref.\,\cite{Kremer:2021} might further improve quantitative accuracy. Nevertheless, the model captures the fundamental observations of the experiment and provides a clear explanation of the underlying mechanisms.   The simulated spectral function in the inset of  Fig.\,\ref{fig:fig4}\textbf{d} showcases the calculated dispersion of QWS1 at the two representative time delays, highlighting the faithful reproduction of the TR-ARPES data. Finally, the PV model can also account for the long timescale (950\,ps) needed to recover equilibrium conditions (Fig.\,\ref{fig:fig2}\textbf{c}), as the spatial separation of electrons and holes in the SCR drastically reduces the recombination rate. The same timescale is expected for the Rashba splitting to recover its initial value as it is modified by the same effect. Since the return to equilibrium is ultimately determined by the charge carriers' diffusion from the illuminated area, the lifetime of the PV effect could in principle be tuned by varying the size of the pump beam.

In conclusion, we demonstrated that light can be used to control the Rashba spin splitting and, by extension, the spin transport properties in semiconductor devices. Specifically, an optically driven photovoltage can be used to manipulate the surface band dispersion and electron distribution at ultrafast (picosecond) timescales. 
The specific application of this technique on 2DEGs to tune the Rashba-splitting on a picosecond timescale is an important benchmark for the development of optically controlled spin devices.
While the implementation of this effect in a working device is by no means trivial, it is informative to contextualize our finding within the framework of the spinFET discussed in Fig.\,\ref{fig:fig1}\textbf{b}: the observed variation of $\Delta k_{R}$ in QWS1, about 3.5$\times 10^{-3}$\,\AA$^{-1}$, translates into a difference in spin precession angle of $\pi$ after $<$100\,nm travel distance, making this effect theoretically appreciable in devices of such length, where ballistic transport can be achieved. It is important to emphasize that, while this study is performed on a TI platform, the underlying physics does not require topological non-triviality and is universal to semiconductors. The effect of the PV on the Rashba strength can be enhanced producing 2DEGs with higher effective masses, while surface gating and pump fluence can be utilized as tuning parameters (see Supplementary Information).

\textbf{Methods:}
Samples of Ca-doped Bi$_{2}$Se$_{3}$ are synthesized as described in Ref.\,\onlinecite{Bianchi:2012}. Here, Ca acts as acceptor atom, positively doping Bi$_{2}$Se$_{3}$ which is normally found to be n-doped due to Se vacancies. The samples are cleaved in vacuum at pressures lower than $7\cdot 10^{-11}$\,mbar, and kept at a temperature of 15\,K during evaporation and measurements. 2DEGs are induced by evaporating K (Fig.\,\ref{fig:fig2}) or Li (Fig.\,\ref{fig:fig3}) \textit{in situ} on the cleaved sample surface. 
The TR-ARPES experiments are performed at QMI's UBC-Moore Center for Ultrafast Quantum Matter \cite{Mills:2019, Boschini:2017}, with 1.55 and 6.2\,eV photons for pump and probe, respectively. Both pump and probe have linear horizontal polarization (parallel to the analyzer slit direction). The pump (probe) beam radius is 150\,$\mu m$ (100\,$\mu m$), and the pump fluence is 40 and 80 $\mu J/cm^{2}$ for experiments represented in Fig.\,\ref{fig:fig2} and \ref{fig:fig3}, respectively. Pump and probe were collinear with an incidence angle of 45\,degrees with respect to the sample normal. Energy and temporal resolution are 17\,meV and 250\,fs, respectively, as determined by the width of the gold Fermi edge and of the combined pump-probe dynamics of the pump induced direct population peak in Bi$_{2}$Se$_{3}$ \cite{Sobota:2014}.
For the band bending model, the Poisson equation was solved numerically employing a modified Thomas-Fermi approximation, which intrinsically accounts for modulation of the charge density due to confinement-induced quantization, without the need for numerically heavy self-consistent calculations \cite{Trott:1993, King:2008}. The Schr\"{o}dinger equation was solved numerically with the Numerov algorithm \cite{Aguiar:2005}. The code is available at Ref.\,\onlinecite{Michiardi:BBcode}.

\textbf{Data Availability:} 
The authors declare that the main data supporting the findings of this study are available within the paper and its Supplementary Information files.
The raw ARPES data generated in this study have been deposited in the Zenodo database under the digital object identifier 10.5281/zenodo.6471678. Extra data are available from the corresponding authors upon request.

\textbf{Acknowledgements:}
We would like to thank M. Bianchi and P. D. C. King for the fruitful discussions. This research was undertaken thanks in part to funding from the Max Planck-UBC-UTokyo Centre for Quantum Materials and the Canada First Research Excellence Fund, Quantum Materials and Future Technologies Program. This project is also funded by the Gordon and Betty Moore Foundation's EPiQS Initiative, Grant GMBF4779 to A.D. and D.J.J.; the Killam, Alfred P. Sloan, and Natural Sciences and Engineering Research Council of Canada's (NSERC's) Steacie Memorial Fellowships (A.D.); the Alexander von Humboldt Foundation (A.D.); the Canada Research Chairs Program (A.D.); NSERC, Canada Foundation for Innovation (CFI); the British Columbia Knowledge Development Fund (BCKDF); the Department of National Defence (DND), the VILLUM FONDEN via the Centre of Excellence for Dirac Materials (Grant No. 11744); and the CIFAR Quantum Materials Program.

\textbf{Author Contributions:}
M.M, F.B, H-H.K, M-X.N performed the measurements and data analysis, S.K.Y.D contributed to the measurements, A.C contributed to data analysis, G.L., S.Z., A.K.M, D.J.J  provided technical support and instrumentation, J.L.M, B.B.I, and P.H. provided the samples, P.H. and A.D. were responsible for the overall direction, planning, and management of the project. All authors contributed to the manuscript. 

\textbf{Competing Interests:}
Authors declare no competing interests.


\begin{thebibliography}{10}
\expandafter\ifx\csname url\endcsname\relax
  \def\url#1{\texttt{#1}}\fi
\expandafter\ifx\csname urlprefix\endcsname\relax\def\urlprefix{URL }\fi
\providecommand{\bibinfo}[2]{#2}
\providecommand{\eprint}[2][]{\url{#2}}

\bibitem{Oestreich:2001}
\bibinfo{author}{Oestreich, M.} \emph{et~al.}
\newblock \bibinfo{title}{{Spintronics: Spin electronics and optoelectronics in
  semiconductors}}.
\newblock \emph{\bibinfo{journal}{Advances in Solid State Physics}}
  \textbf{\bibinfo{volume}{41}}, \bibinfo{pages}{173--186}
  (\bibinfo{year}{2001}).
  
\bibitem{Bychkov:1984}
\bibinfo{author}{Bychkov, Y.~A.} \& \bibinfo{author}{Rashba, E.~I.}
\newblock \bibinfo{title}{{Properties of a 2D electron gas with a lifted
  spectrum degeneracy}}.
\newblock \emph{\bibinfo{journal}{Journal of Experimental and Theoretical
  Physics Letters}} \bibinfo{pages}{78--81} (\bibinfo{year}{1984}).

\bibitem{Rashba:2006}
\bibinfo{author}{Rashba, E.}
\newblock \bibinfo{title}{{Spin–orbit coupling and spin transport}}.
\newblock \emph{\bibinfo{journal}{Physica E: Low-dimensional Systems and
  Nanostructures}} \textbf{\bibinfo{volume}{34}}, \bibinfo{pages}{31--35}
  (\bibinfo{year}{2006}).

\bibitem{Bercioux:2015}
\bibinfo{author}{Bercioux, D.} \& \bibinfo{author}{Lucignano, P.}
\newblock \bibinfo{title}{{Quantum transport in Rashba spin-orbit materials: a
  review}}.
\newblock \emph{\bibinfo{journal}{Reports on Progress in Physics}}
  \textbf{\bibinfo{volume}{78}}, \bibinfo{pages}{106001}
  (\bibinfo{year}{2015}).

\bibitem{Soumyanarayanan:2016}
\bibinfo{author}{Soumyanarayanan, A.}, \bibinfo{author}{Reyren, N.},
  \bibinfo{author}{Fert, A.} \& \bibinfo{author}{Panagopoulos, C.}
\newblock \bibinfo{title}{{Emergent phenomena induced by spin–orbit coupling
  at surfaces and interfaces}}.
\newblock \emph{\bibinfo{journal}{Nature}} \textbf{\bibinfo{volume}{539}},
  \bibinfo{pages}{509--517} (\bibinfo{year}{2016}).
  
\bibitem{Datta:1990}
\bibinfo{author}{Datta, S.} \& \bibinfo{author}{Das, B.}
\newblock \bibinfo{title}{{Electronic analog of the electro‐optic
  modulator}}.
\newblock \emph{\bibinfo{journal}{Applied Physics Letters}}
  \textbf{\bibinfo{volume}{56}}, \bibinfo{pages}{665--667}
  (\bibinfo{year}{1990}).
  
\bibitem{Chuang:2015}
\bibinfo{author}{Chuang, P.} \emph{et~al.}
\newblock \bibinfo{title}{{All-electric all-semiconductor spin field-effect
  transistors}}.
\newblock \emph{\bibinfo{journal}{Nature Nanotechnology}}
  \textbf{\bibinfo{volume}{10}}, \bibinfo{pages}{35--39}
  (\bibinfo{year}{2015}).
  
\bibitem{Yuan:2014}
\bibinfo{author}{Yuan, H.} \emph{et~al.}
\newblock \bibinfo{title}{{Generation and electric control of
  spin–valley-coupled circular photogalvanic current in WSe$_{2}$}}.
\newblock \emph{\bibinfo{journal}{Nature Nanotechnology}}
  \textbf{\bibinfo{volume}{9}}, \bibinfo{pages}{851--857}
  (\bibinfo{year}{2014}).

\bibitem{Liu:2020}
\bibinfo{author}{Liu, X.} \emph{et~al.}
\newblock \bibinfo{title}{{Circular photogalvanic spectroscopy of Rashba
  splitting in 2D hybrid organic–inorganic perovskite multiple quantum
  wells}}.
\newblock \emph{\bibinfo{journal}{Nature Communications}}
  \textbf{\bibinfo{volume}{11}}, \bibinfo{pages}{323} (\bibinfo{year}{2020}).

\bibitem{Kimel:2004}
\bibinfo{author}{Kimel, A.~V.}, \bibinfo{author}{Kirilyuk, A.},
  \bibinfo{author}{Tsvetkov, A.}, \bibinfo{author}{Pisarev, R.~V.} \&
  \bibinfo{author}{Rasing, T.}
\newblock \bibinfo{title}{{Laser-induced ultrafast spin reorientation in the
  antiferromagnet TmFeO$_{3}$}}.
\newblock \emph{\bibinfo{journal}{Nature}} \textbf{\bibinfo{volume}{429}},
  \bibinfo{pages}{850--853} (\bibinfo{year}{2004}).

\bibitem{Nova:2016}
\bibinfo{author}{Nova, T.~F.} \emph{et~al.}
\newblock \bibinfo{title}{{An effective magnetic field from optically driven
  phonons}}.
\newblock \emph{\bibinfo{journal}{Nature Physics}}
  \textbf{\bibinfo{volume}{13}}, \bibinfo{pages}{132--136}
  (\bibinfo{year}{2017}).
  
\bibitem{McIver:2009}
\bibinfo{author}{McIver, J.~W.}, \bibinfo{author}{Hsieh, D.},
  \bibinfo{author}{Steinberg, H.}, \bibinfo{author}{Jarillo-Herrero, P.} \&
  \bibinfo{author}{Gedik, N.}
\newblock \bibinfo{title}{{Control over topological insulator photocurrents
  with light polarization}}.
\newblock \emph{\bibinfo{journal}{Nature Nanotechnology}}
  \textbf{\bibinfo{volume}{7}}, \bibinfo{pages}{96--100}
  (\bibinfo{year}{2012}).

\bibitem{Wang:2007}
\bibinfo{author}{Wang, J.}, \bibinfo{author}{Zhu, B.-F.} \&
  \bibinfo{author}{Liu, R.-B.}
\newblock \bibinfo{title}{{Proposal for direct measurement of a pure spin
  current by a polarized light beam}}.
\newblock \emph{\bibinfo{journal}{Physical Review Letters}}
  \textbf{\bibinfo{volume}{100}}, \bibinfo{pages}{086603}
  (\bibinfo{year}{2007}).

\bibitem{Zhou:2006}
\bibinfo{author}{Zhou, B.} \& \bibinfo{author}{Shen, S.-Q.}
\newblock \bibinfo{title}{{Deduction of pure spin current from the linear and
  circular spin photogalvanic effect in semiconductor quantum wells}}.
\newblock \emph{\bibinfo{journal}{Physical Review B}}
  \textbf{\bibinfo{volume}{75}}, \bibinfo{pages}{045339}
  (\bibinfo{year}{2007}).
  
\bibitem{Sheremet:2016}
\bibinfo{author}{Sheremet, A.~S.}, \bibinfo{author}{Kibis, O.~V.},
  \bibinfo{author}{Kavokin, A.~V.} \& \bibinfo{author}{Shelykh, I.~A.}
\newblock \bibinfo{title}{{Datta-and-Das spin transistor controlled by a
  high-frequency electromagnetic field}}.
\newblock \emph{\bibinfo{journal}{Physical Review B}}
  \textbf{\bibinfo{volume}{93}}, \bibinfo{pages}{165307}
  (\bibinfo{year}{2016}).

\bibitem{Cheng:2017}
\bibinfo{author}{Cheng, L.} \emph{et~al.}
\newblock \bibinfo{title}{{Optical manipulation of Rashba spin–orbit coupling
  at SrTiO$_{3}$-based oxide interfaces}}.
\newblock \emph{\bibinfo{journal}{Nano Letters}} \textbf{\bibinfo{volume}{17}},
  \bibinfo{pages}{6534--6539} (\bibinfo{year}{2017}).

\bibitem{Bianchi:2010}
\bibinfo{author}{Bianchi, M.} \emph{et~al.}
\newblock \bibinfo{title}{{Coexistence of the topological state and a
  two-dimensional electron gas on the surface of Bi$_{2}$Se$_{3}$}}.
\newblock \emph{\bibinfo{journal}{Nature Communications}}
  \textbf{\bibinfo{volume}{1}}, \bibinfo{pages}{128} (\bibinfo{year}{2010}).

\bibitem{Bahramy:2012}
\bibinfo{author}{Bahramy, M.} \emph{et~al.}
\newblock \bibinfo{title}{{Emergent quantum confinement at topological
  insulator surfaces}}.
\newblock \emph{\bibinfo{journal}{Nature Communications}}
  \textbf{\bibinfo{volume}{3}}, \bibinfo{pages}{1159} (\bibinfo{year}{2012}).

\bibitem{Zhu:2011}
\bibinfo{author}{Zhu, Z.-H.} \emph{et~al.}
\newblock \bibinfo{title}{{Rashba spin-splitting control at the surface of the
  topological insulator Bi$_{2}$Se$_{3}$}}.
\newblock \emph{\bibinfo{journal}{Physical Review Letters}}
  \textbf{\bibinfo{volume}{107}}, \bibinfo{pages}{186405}
  (\bibinfo{year}{2011}).

\bibitem{Michiardi:2015}
\bibinfo{author}{Michiardi, M.} \emph{et~al.}
\newblock \bibinfo{title}{{Strongly anisotropic spin-orbit splitting in a
  two-dimensional electron gas}}.
\newblock \emph{\bibinfo{journal}{Physical Review B}}
  \textbf{\bibinfo{volume}{91}}, \bibinfo{pages}{035445}
  (\bibinfo{year}{2015}).

\bibitem{Barreto:2013}
\bibinfo{author}{Barreto, L.} \emph{et~al.}
\newblock \bibinfo{title}{{Surface-dominated transport on a bulk topological
  insulator}}.
\newblock \emph{\bibinfo{journal}{Nano Letters}} \textbf{\bibinfo{volume}{14}},
  \bibinfo{pages}{3755--3760} (\bibinfo{year}{2014}).

\bibitem{Perfetti:2006}
\bibinfo{author}{Perfetti, L.} \emph{et~al.}
\newblock \bibinfo{title}{{Time evolution of the electronic structure of
  1T-TaS$_{2}$ through the insulator-metal transition}}.
\newblock \emph{\bibinfo{journal}{Physical Review Letters}}
  \textbf{\bibinfo{volume}{97}}, \bibinfo{pages}{067402}
  (\bibinfo{year}{2006}).

\bibitem{Freericks:2008}
\bibinfo{author}{Freericks, J.~K.}, \bibinfo{author}{Krishnamurthy, H.~R.} \&
  \bibinfo{author}{Pruschke, T.}
\newblock \bibinfo{title}{{Theoretical description of time-resolved
  photoemission spectroscopy: Application to pump-probe experiments}}.
\newblock \emph{\bibinfo{journal}{Physical Review Letters}}
  \textbf{\bibinfo{volume}{102}}, \bibinfo{pages}{136401}
  (\bibinfo{year}{2009}).

\bibitem{Sobota:2014}
\bibinfo{author}{Sobota, J.} \emph{et~al.}
\newblock \bibinfo{title}{{Ultrafast electron dynamics in the topological
  insulator Bi$_{2}$Se$_{3}$ studied by time-resolved photoemission
  spectroscopy}}.
\newblock \emph{\bibinfo{journal}{Journal of Electron Spectroscopy and Related
  Phenomena}} \textbf{\bibinfo{volume}{195}}, \bibinfo{pages}{249--257}
  (\bibinfo{year}{2014}).

\bibitem{Hajlaoui:2012}
\bibinfo{author}{Hajlaoui, M.} \emph{et~al.}
\newblock \bibinfo{title}{{Ultrafast surface carrier dynamics in the
  topological insulator Bi$_{2}$Te$_{3}$}}.
\newblock \emph{\bibinfo{journal}{Nano Letters}} \textbf{\bibinfo{volume}{12}},
  \bibinfo{pages}{3532--3536} (\bibinfo{year}{2012}).

\bibitem{Ciocys:2019}
\bibinfo{author}{Ciocys, S.}, \bibinfo{author}{Morimoto, T.},
  \bibinfo{author}{Moore, J.~E.} \& \bibinfo{author}{Lanzara, A.}
\newblock \bibinfo{title}{{Tracking surface photovoltage dipole geometry in
  Bi$_{2}$Se$_{3}$ with time-resolved photoemission}}.
\newblock \emph{\bibinfo{journal}{Journal of Statistical Mechanics: Theory and
  Experiment}} \textbf{\bibinfo{volume}{2019}}, \bibinfo{pages}{104008}
  (\bibinfo{year}{2019}).

\bibitem{Yang:2013}
\bibinfo{author}{Yang, S.-L.}, \bibinfo{author}{Sobota, J.~A.},
  \bibinfo{author}{Kirchmann, P.~S.} \& \bibinfo{author}{Shen, Z.-X.}
\newblock \bibinfo{title}{{Electron propagation from a photo-excited surface:
  implications for time-resolved photoemission}}.
\newblock \emph{\bibinfo{journal}{Applied Physics A}}
  \textbf{\bibinfo{volume}{116}}, \bibinfo{pages}{85--90}
  (\bibinfo{year}{2014}).

\bibitem{Tanaka:2012}
\bibinfo{author}{Tanaka, S.-i.}
\newblock \bibinfo{title}{{Utility and constraint on the use of pump-probe
  photoelectron spectroscopy for detecting time-resolved surface
  photovoltage}}.
\newblock \emph{\bibinfo{journal}{Journal of Electron Spectroscopy and Related
  Phenomena}} \textbf{\bibinfo{volume}{185}}, \bibinfo{pages}{152--158}
  (\bibinfo{year}{2012}).

\bibitem{Paasch:1982}
\bibinfo{author}{Paasch, G.} \& \bibinfo{author}{\"Ubensee, H.}
\newblock \bibinfo{title}{{A Modified Local Density Approximation. Electron
  Density in Inversion Layers}}.
\newblock \emph{\bibinfo{journal}{Physica Status Solidi (b)}}
  \textbf{\bibinfo{volume}{113}}, \bibinfo{pages}{165--178}
  (\bibinfo{year}{1982}).

\bibitem{King:2008}
\bibinfo{author}{King, P. D.~C.}, \bibinfo{author}{Veal, T.~D.} \&
  \bibinfo{author}{McConville, C.~F.}
\newblock \bibinfo{title}{{Nonparabolic coupled Poisson-Schrödinger solutions
  for quantized electron accumulation layers: Band bending, charge profile, and
  subbands at InN surfaces}}.
\newblock \emph{\bibinfo{journal}{Physical Review B}}
  \textbf{\bibinfo{volume}{77}}, \bibinfo{pages}{125305}
  (\bibinfo{year}{2008}).

\bibitem{Analytis:2010}
\bibinfo{author}{Analytis, J.~G.} \emph{et~al.}
\newblock \bibinfo{title}{{Bulk Fermi surface coexistence with Dirac surface
  state in Bi$_{2}$Se$_{3}$: A comparison of photoemission and Shubnikov–de
  Haas measurements}}.
\newblock \emph{\bibinfo{journal}{Physical Review B}}
  \textbf{\bibinfo{volume}{81}}, \bibinfo{pages}{205407}
  (\bibinfo{year}{2010}).

\bibitem{Martinez:2017}
\bibinfo{author}{Martinez, G.} \emph{et~al.}
\newblock \bibinfo{title}{{Determination of the energy band gap of
  Bi$_{2}$Se$_{3}$}}.
\newblock \emph{\bibinfo{journal}{Scientific Reports}}
  \textbf{\bibinfo{volume}{7}}, \bibinfo{pages}{6891} (\bibinfo{year}{2017}).

\bibitem{Nechaev:2013}
\bibinfo{author}{Nechaev, I.~A.} \emph{et~al.}
\newblock \bibinfo{title}{{Evidence for a direct band gap in the topological
  insulator Bi$_{2}$Se$_{3}$ from theory and experiment}}.
\newblock \emph{\bibinfo{journal}{Physical Review B}}
  \textbf{\bibinfo{volume}{87}}, \bibinfo{pages}{121111}
  (\bibinfo{year}{2013}).

\bibitem{Gao:2014}
\bibinfo{author}{Gao, Y.-B.}, \bibinfo{author}{He, B.},
  \bibinfo{author}{Parker, D.}, \bibinfo{author}{Androulakis, I.} \&
  \bibinfo{author}{Heremans, J.~P.}
\newblock \bibinfo{title}{{Experimental study of the valence band of
  Bi$_{2}$Se$_{3}$}}.
\newblock \emph{\bibinfo{journal}{Physical Review B}}
  \textbf{\bibinfo{volume}{90}}, \bibinfo{pages}{125204}
  (\bibinfo{year}{2014}).

\bibitem{Kronik:1999}
\bibinfo{author}{Kronik, L.} \& \bibinfo{author}{Shapira, Y.}
\newblock \bibinfo{title}{{Surface photovoltage phenomena: theory, experiment,
  and applications}}.
\newblock \emph{\bibinfo{journal}{Surface Science Reports}}
  \textbf{\bibinfo{volume}{37}}, \bibinfo{pages}{1--206}
  (\bibinfo{year}{1999}).

\bibitem{Yoshikawa:2019}
\bibinfo{author}{Yoshikawa, T.} \emph{et~al.}
\newblock \bibinfo{title}{{Bidirectional surface photovoltage on a topological
  insulator}}.
\newblock \emph{\bibinfo{journal}{Physical Review B}}
  \textbf{\bibinfo{volume}{100}}, \bibinfo{pages}{165311}
  (\bibinfo{year}{2019}).

\bibitem{SanchezBarriga:2017}
\bibinfo{author}{Sánchez-Barriga, J.} \emph{et~al.}
\newblock \bibinfo{title}{{Laser-induced persistent photovoltage on the surface
  of a ternary topological insulator at room temperature}}.
\newblock \emph{\bibinfo{journal}{Applied Physics Letters}}
  \textbf{\bibinfo{volume}{110}}, \bibinfo{pages}{141605}
  (\bibinfo{year}{2017}).

\bibitem{Ciocys:2020}
\bibinfo{author}{Ciocys, S.} \emph{et~al.}
\newblock \bibinfo{title}{{Manipulating long-lived topological surface
  photovoltage in bulk-insulating topological insulators Bi$_{2}$Se$_{3}$ and
  Bi$_{2}$Te$_{3}$}}.
\newblock \emph{\bibinfo{journal}{npj Quantum Materials}}
  \textbf{\bibinfo{volume}{5}}, \bibinfo{pages}{16} (\bibinfo{year}{2020}).

\bibitem{MoraSero:2006}
\bibinfo{author}{Mora-Seró, I.}, \bibinfo{author}{Dittrich, T.},
  \bibinfo{author}{Garcia-Belmonte, G.} \& \bibinfo{author}{Bisquert, J.}
\newblock \bibinfo{title}{{Determination of spatial charge separation of
  diffusing electrons by transient photovoltage measurements}}.
\newblock \emph{\bibinfo{journal}{Journal of Applied Physics}}
  \textbf{\bibinfo{volume}{100}}, \bibinfo{pages}{103705}
  (\bibinfo{year}{2006}).

\bibitem{Kremer:2021}
\bibinfo{author}{Kremer, G.} \emph{et~al.}
\newblock \bibinfo{title}{Ultrafast dynamics of the surface photovoltage in
  potassium-doped black phosphorus}.
\newblock \emph{\bibinfo{journal}{Physical Review B}}
  \textbf{\bibinfo{volume}{104}}, \bibinfo{pages}{035125}
  (\bibinfo{year}{2021}).

\bibitem{Bianchi:2012}
\bibinfo{author}{Bianchi, M.} \emph{et~al.}
\newblock \bibinfo{title}{{Robust surface doping of Bi$_{2}$Se$_{3}$ by
  rubidium intercalation}}.
\newblock \emph{\bibinfo{journal}{ACS Nano}} \textbf{\bibinfo{volume}{6}},
  \bibinfo{pages}{7009--7015} (\bibinfo{year}{2012}).

\bibitem{Mills:2019}
\bibinfo{author}{Mills, A.~K.} \emph{et~al.}
\newblock \bibinfo{title}{{Cavity-enhanced high harmonic generation for extreme
  ultraviolet time- and angle-resolved photoemission spectroscopy}}.
\newblock \emph{\bibinfo{journal}{Review of Scientific Instruments}}
  \textbf{\bibinfo{volume}{90}}, \bibinfo{pages}{083001}
  (\bibinfo{year}{2019}).

\bibitem{Boschini:2017}
\bibinfo{author}{Boschini, F.} \emph{et~al.}
\newblock \bibinfo{title}{{Collapse of superconductivity in cuprates via
  ultrafast quenching of phase coherence}}.
\newblock \emph{\bibinfo{journal}{Nature Materials}}
  \textbf{\bibinfo{volume}{17}}, \bibinfo{pages}{416--420}
  (\bibinfo{year}{2018}).

\bibitem{Trott:1993}
\bibinfo{author}{Trott, S.}, \bibinfo{author}{Trott, M.} \&
  \bibinfo{author}{Nakov, V.}
\newblock \bibinfo{title}{{Modified Thomas‐Fermi Approximation. A
  Surprisingly Good Tool for the Treatment of Semiconductor Layer Structures
  Including Various Two‐Dimensional Systems}}.
\newblock \emph{\bibinfo{journal}{Phys Status Solidi B}}
  \textbf{\bibinfo{volume}{177}}, \bibinfo{pages}{389--395}
  (\bibinfo{year}{1993}).

\bibitem{Aguiar:2005}
\bibinfo{author}{Vigo-Aguiar, J.} \& \bibinfo{author}{Ramos, H.}
\newblock \bibinfo{title}{{A variable-step Numerov method for the numerical
  solution of the Schr\"{o}dinger equation}}.
\newblock \emph{\bibinfo{journal}{Journal of Mathematical Chemistry}}
  \textbf{\bibinfo{volume}{37}}, \bibinfo{pages}{255--262}
  (\bibinfo{year}{2005}).

\bibitem{Michiardi:BBcode}
\bibinfo{author}{Michiardi, M.}
\newblock \bibinfo{title}{{P-MTFA-S} band bending simulation}
  (\bibinfo{year}{2021}).
\newblock
  \urlprefix\url{https://github.com/okio-mm/P-MTFA-S_band_bending_simulation}.

\bibitem{Yang:2008}
\bibinfo{author}{Yang, H.-B.} \emph{et~al.}
\newblock \bibinfo{title}{{Emergence of preformed Cooper pairs from the doped
  Mott insulating state in Bi$_{2}$Sr$_{2}$CaCu$_{2}$O$_{8+\delta}$}}.
\newblock \emph{\bibinfo{journal}{Nature}} \textbf{\bibinfo{volume}{456}},
  \bibinfo{pages}{77--80} (\bibinfo{year}{2008}).

\end{thebibliography}

\newpage

\end{document}